\documentclass[10pt,twocolumn]{article}
\usepackage[T1]{fontenc}
\usepackage{times}
\usepackage[margin=0.75in]{geometry}
\usepackage{booktabs}
\usepackage{graphicx}
\graphicspath{{./}}
\usepackage{amssymb}
\usepackage{amsmath}
\usepackage[numbers]{natbib}   
\usepackage[hidelinks]{hyperref}
\usepackage{balance}   
\title{\textbf{Cultural Targets, Structural Frames, Binding Morals:\\
A Cross-Lingual Audit of Online Hate in Multicultural Singapore}}
\author{Emilio Ferrara\\
  \small Thomas Lord Department of Computer Science, University of Southern California\\
  \small Centre for Advanced Technologies in Online Safety (CATOS), Institute of High Performance Computing (IHPC), A*STAR, Singapore}
\date{}
\begin{document}
\maketitle

\begin{abstract}
\noindent Multicultural Singapore hosts overlapping language publics (English, Chinese, and Malay) that
discuss the same out-groups in parallel, a natural setting to ask whether online hate shares a structure
across languages and whether what a community \emph{produces} is what it \emph{amplifies}. From a
Singapore-centric 2025 Facebook, Reddit, and YouTube corpus (31.0M items; 1.76M comments mentioning eleven
identity groups), we benchmark eight open large language models as hate annotators against a
human-adjudicated gold set, adopt the best (Phi-4: accuracy 0.95, Cohen's $\kappa$=0.91, recall 1.00 on an
independent manual check), and replicate every finding under a second model. The results converge on one
thesis, \emph{layered cultural contingency}: cross-lingual divergence falls monotonically as one moves from
what a community hates to how and why it hates. Which out-groups are targeted is culturally specific
(language $\times$ target $V$=0.25), but the threat frames and the binding moral grammar of hate (sanctity
and loyalty, 55--75\%, not fairness) are far more shared across languages, with divergence dropping to
$V$=0.08 for moral foundations and 0.07 for emotion. Hate is contempt-driven and voices an out-group,
anti-immigration grievance rather than an anti-system one. Reception is selectively nativist: hateful
comments are amplified less than neutral mentions overall, yet anti-immigrant hate is preferentially
amplified while religious and anti-LGBTQ hate is not, and volume does not track 2025 Singapore key events. We further show
that absolute hate prevalence is not well defined at the LLM-annotator level, with agreement ceilings at
$\kappa\approx0.42$ across models, so we report relative structure as primary. The findings bear directly on
cross-lingual content moderation.
\end{abstract}

\section{Introduction}
Online hate is usually studied one language and one platform at a time, almost always from the sender's side:
what makes a message hateful, and how to detect it. Two questions are less understood. First, whether the
\emph{structure} of hate (which groups are targeted, how the hostility is framed, what moral language carries
it) is shared across the language publics that coexist within one society, or whether each public hates in its
own idiom. Second, whether the hate a community \emph{produces} is the hate its audience \emph{amplifies}.
Both questions bear on moderation: a platform serving a multilingual population needs to know whether a
detector or policy calibrated in one language transfers to another, and whether reducing the production of
hate is the same goal as reducing its reach.

Multicultural Singapore is a clean setting in which to ask them. Its population debates a shared set of
out-groups (local ethnic communities under the state's Chinese--Malay--Indian--Other (CMIO) framework,
mainland-Chinese (``PRC'') nationals, Indian professionals admitted under the CECA agreement, migrant workers,
religious groups, and sexual minorities) simultaneously in English, Chinese, and Malay. The same events
and tensions surface in parallel across these publics within a single year, holding social context roughly
fixed while the language of expression varies.

We assemble a Singapore-centric Facebook, Reddit, and YouTube corpus spanning 2025 (31.0M items across 252
spaces; Table~\ref{tab:corpus}) and identify 1.76M comments mentioning eleven Singapore-salient out-groups.
Measuring hate across three languages at scale surfaces a methodological problem: LLM annotators, now a default
labeling instrument \citep{gilardi2023chatgpt}, disagree with one another on how much hate exists. Across
eight open models scored against a 211-item human gold set, agreement ceilings at Cohen's $\kappa\approx0.42$
and is model-dependent. Model \emph{choice} dominates \emph{size} (a 14B model beats a 72B), and one family
collapses on Chinese. We therefore treat absolute hate prevalence as ill-defined at the annotator level and
report relative structure as primary, absolute rates only as calibrated ranges. We adopt the best-validated
model, Phi-4, as the labeler of record (accuracy 0.95, $\kappa$=0.91, recall 1.00 against an independent manual
check of the labels in use), and replicate every finding under a second, independent model so that no
conclusion rests on a single annotator. The validated pipeline yields 3{,}323 confirmed-hate comments
(English 1{,}584; Chinese 1{,}336; Malay 403).

\begin{figure*}[t]\centering
\includegraphics[width=\textwidth]{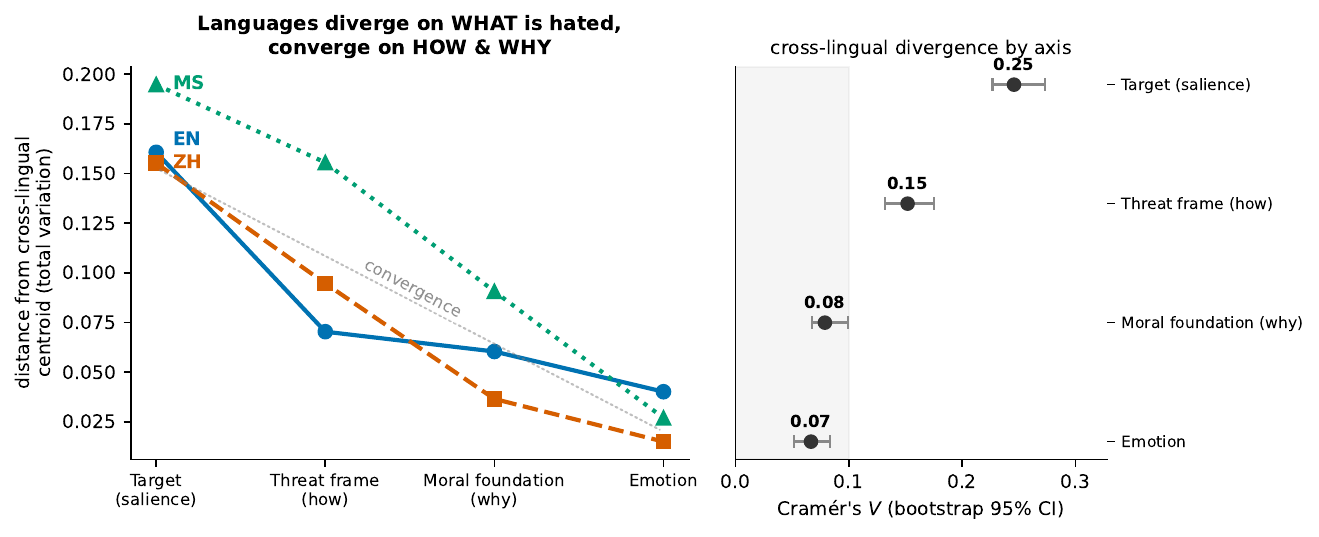}
\caption{The layered cultural contingency of online hate. \textbf{Left:} each language's distribution
distance (total variation) from the cross-lingual centroid, per axis. The three language publics start far
apart on \emph{which} group is hated (target salience) and \emph{funnel toward convergence} on \emph{how}
and \emph{why} (frame, moral foundation, emotion). \textbf{Right:} cross-lingual divergence as Cram\'er's
$V$ with bootstrap 95\% CIs, declining monotonically and significantly from target ($V$=0.25 [0.22, 0.27])
to frame (0.15 [0.13, 0.18]) to moral foundation (0.08) and emotion (0.07). The \emph{what} of hate is
local, while its \emph{grammar} is cross-lingually shared.}
\label{fig:hero}
\end{figure*}

With that discipline, the findings cohere into a single structural claim rather than four separate
observations: \textbf{online hate is layered by cultural contingency.} Its \emph{targets} are culturally
selected and vary sharply across language publics, but its \emph{grammar} (the threat frame it reaches
for and the moral foundation it invokes) is cross-lingually near-invariant, and the divergence falls
monotonically as one moves from what is hated to how and why it is hated (Figure~\ref{fig:hero}). Concretely:
\textbf{what a community hates is cultural, how it frames hate is structural, how it moralizes hate is
near-universal and binding-based, and what it amplifies is selectively nativist.} This layered view is
falsifiable and is exactly what our cross-lingual, single-society design is positioned to test. The corpus
is \emph{actor-less} (no user identities), so every claim is about comments, not users.

We organize the analysis around three research questions, in logical order (structure, then expression, then
reception):
\begin{itemize}\setlength{\itemsep}{2pt}
\item \textbf{RQ1 (structure).} Is online hate culturally contingent or cross-lingually shared, and at which
layers: (a) targets, (b) threat frames, (c) moral foundations?
\item \textbf{RQ2 (expression).} How is hate substantively expressed in this multicultural setting:
(a) Singapore narratives, (b) affect and dehumanization, (c) political stance?
\item \textbf{RQ3 (reception).} Does the audience amplify what is produced: (a) production versus resonance,
(b) temporal event-reactivity?
\end{itemize}
RQ1 tests the layered-contingency thesis (Figure~\ref{fig:hero}), RQ2 grounds it in local content, and RQ3 turns
to the under-studied reception side.

\paragraph{Contributions of this paper.} (i) The \emph{layered-contingency} account of online hate: a
falsifiable claim that the cultural variability of hate decreases monotonically from target to frame to moral
foundation (Figure~\ref{fig:hero}), evidenced across three language publics and replicated under two LLM
labelers. (ii) The production-versus-resonance dissociation, foregrounding the under-studied reception side of
hate. (iii) A measurement-validity result of independent interest: absolute hate prevalence is not well
defined at the annotator level, model choice beats size, and English-centric single-family annotation
degrades on other languages.

\section{Related Work}
\paragraph{Identity, threat, and the form of intergroup hate.} Social Identity Theory grounds out-group
derogation in the universal in-group/out-group categorization process, while making the \emph{content} of the
salient out-group context-dependent \citep{tajfel1979integrative}, the substrate of our claim that the
\emph{what} of hate is locally selected but its \emph{form} is shared. Integrated Threat Theory distinguishes
\emph{realistic} from \emph{symbolic} threat \citep{stephan2000itt,stephan2016itt}. Intergroup-emotions theory
ties specific out-group emotions (contempt, anger, disgust) to action tendencies \citep{mackie2000intergroup}.
And dehumanization \citep{haslam2006dehumanization}, moral exclusion \citep{opotow1990moral}, and the
disgust$\rightarrow$dehumanization$\rightarrow$prejudice pathway \citep{hodson2007interpersonal} describe
\emph{cross-culturally recurrent} mechanisms once a target is selected. We operationalize a six-way
threat-frame taxonomy and test whether the frame is target-determined and cross-lingually invariant. Immigration
and ``great replacement'' framing have established political-attitudes \citep{herold2025replacement} and
computational \citep{mendelsohn2021framing} literatures.

\paragraph{Moral foundations of hate.} Moral Foundations Theory \citep{graham2009mft,graham2011mapping}
distinguishes \emph{individualizing} (care, fairness) from \emph{binding} (loyalty, authority, sanctity)
foundations. The binding foundations are the ones empirically tied to out-group hostility and
authoritarianism \citep{kugler2014another}, and they mediate anti-migrant prejudice via ITT-style threat
\citep{bianco2023binding}. Moralized language predicts hate at scale \citep{solovev2022moralized}, and moral
framing shapes online communication \citep{jiang2025moral} \citep[dictionary tools:][]{hopp2021emfd}.
These foundations appear \emph{universally} but are \emph{weighted} differently across cultures
\citep{atari2020iran}, the prototype for our ``universal layer, variable emphasis'' claim. We find binding
foundations dominate hate cross-lingually and report \emph{within-language} comparisons given the labeling
ceiling.

\paragraph{Measuring online hate across languages.} Population-scale work maps hate as a self-organizing
cross-platform ecology \citep{johnson2019hate}. Detection foundations separate hate from mere offensiveness
\citep{davidson2017automated} and surface implicit, framing-carried hate \citep{elsherief2021latent}.
English-centric detectors transfer poorly across languages \citep{rottger2021hatecheck,rottger2022multihatecheck},
a gap that widens in lower-resource settings where models behave unevenly \citep{haider2023lowresource},
motivating region-specific resources (SGHateCheck \citep{ng2024sghatecheck}, which covers Singapore Malay, COLD and ToxiCN for Chinese
\citep{deng2022cold,lu2023toxicn}, Indonesian abusive-language data \citep{ibrohim2019indonesian} as a
close cross-lingual proxy for Malay, and
East-Asian-prejudice detection \citep{vidgen2020eastasian}) which we use for high-recall retrieval. Because
labels for a contested construct are annotator-dependent \citep{sap2022annotators,davani2022disagreements} and
LLM annotators disagree \citep{gilardi2023chatgpt} and carry cross-cultural skew \citep{dey2025cultural}, we
treat absolute prevalence cautiously. Our contribution
is the structural comparison, not a detector.

\paragraph{Amplification, resonance, and counter-speech.} Moral-emotional language travels farther
\citep{brady2017contagion} and social feedback amplifies it \citep{brady2021amplify}. Out-group animosity in
particular drives engagement \citep{rathje2021outgroup} and online toxicity is shaped by social approval,
network homophily \citep{jiang2024toxicity}, and the moderation regime of the space
\citep{lerman2025moderation}, the reception mechanisms our RQ3 speaks to. Counter-speech can
intervene \citep{mathew2019counterspeech,garland2022counterspeech,hangartner2021empathy}. Most hate research
stops at production. We address reception directly.

\section{Data}
\paragraph{Source and collection.} The corpus is drawn from a social-listening platform (Resonance) that
continuously crawls a \emph{curated list} of Singapore-centric public sources, not a keyword sample of the
open web. The source list comprises 231 Facebook pages (224 Singapore-based, 7 international, predominantly
news, community, and alternative-news pages), 23 subreddits (Singapore community subs such as
\texttt{r/singapore} and \texttt{r/askSingapore}, \texttt{r/malaysia}, general \texttt{r/worldnews} and
\texttt{r/conspiracy}, and a cluster of infectious-disease subreddits reflecting the platform's original
public-health-listening focus), and 16 YouTube channels (Singapore mainstream outlets such as Channel NewsAsia,
The Straits Times, Mothership, and TODAY, plus alternative-news and creators). Facebook and Reddit have been crawled
since July 2020 and YouTube since December 2023, with pages added over time. We analyze the 2025 slice.
Pages are labeled by a primary language (almost all English), but their \emph{comment streams} are
multilingual, which provides the Chinese and Malay content we study.

\paragraph{Composition.} The 2025 slice is a Singapore-centric corpus across Facebook, Reddit, and YouTube:
31.0M items across 252 spaces (Table~\ref{tab:corpus}), overwhelmingly comment-level and trilingual
(English 28.6M, Simplified Chinese 1.62M, Malay 0.79M). Chinese and Malay appear almost
entirely on Facebook, while Reddit and YouTube are English-dominant. Each item carries platform, thread structure,
timestamp, an engagement count, the original text, and a machine-English translation used as the common
labeling surface. A lexical detector for eleven Singapore-salient out-groups (three local races, four
nationality/migrant categories including PRC and CECA, two religions, LGBTQ, and migrant workers) identifies
1.76M target-mentioning comments (English 1.53M, Chinese 182k, Malay 44k). The corpus is actor-less: it
carries no user identifiers, so the analysis is structurally incapable of profiling individuals.

\begin{table}[tb]\centering\small
\caption{Corpus composition (Singapore-centric Facebook/Reddit/YouTube, 2025).}
\label{tab:corpus}
\resizebox{\columnwidth}{!}{%
\begin{tabular}{llr}
\toprule
Platform & Type & Count \\
\midrule
Facebook & comments & 26{,}153{,}326 \\
Facebook & posts & 263{,}084 \\
Reddit & comments & 3{,}548{,}594 \\
YouTube & comments & 796{,}320 \\
\midrule
All & (EN 28.6M / ZH 1.62M / MS 0.79M) & 31{,}032{,}005 \\
\bottomrule
\end{tabular}}
\end{table}

\section{Methods}
A single prompt that judges hate together with its frame and moral content primes the model toward ``hate''
and inflates prevalence, so we separate detection from characterization and validate detection against human
labels.

\paragraph{Two-stage classifier.} Stage~1 is a high-recall lexical and prototype filter (drawing on the
resources above plus provider toxicity signals), producing a candidate pool. Stage~2 is an isolated
\emph{binary} hate decision whose prompt makes the nulls explicit (neutral mention, questions, in-group talk,
criticism of policies or individuals, and counter-speech are not hate), with ties resolved to ``not hate'' and
no frame or moral options present. Confirmation requires consensus across two independently worded prompts.
Confirmed hate is characterized in a separate call for threat frame, moral foundations, Singapore narrative,
emotion, dehumanization, and political stance.

\paragraph{Labeler benchmark and validation.} We benchmarked eight open models' Stage-2 gates against a
211-item human-adjudicated gold set drawn from \emph{candidate-hate} (a deliberately hard, ambiguity-rich
slice; Table~\ref{tab:bench}). On this hard set agreement ceilings at $\kappa\approx0.42$ across all eight
models (no model, regardless of size, separates hate from non-hate cleanly on contested cases), so \emph{absolute} prevalence
is not well defined at the annotator level. We stress this is a property of the hard set: on a separate
random-stratified sample of the labels actually deployed in the corpus (human $n$=89), the best model agrees
with the human far more strongly (below). These two $\kappa$ measure different things, namely inter-annotator
reliability on ambiguous boundary cases (0.42) versus accuracy of the deployed labels on a corpus-representative
sample (0.91), and we report both rather than choosing the flattering one. A model-choice effect dominates
size: a 14B model (Phi-4) beats a 72B model. We note the 72B ran at 4-bit quantization, so this is suggestive,
not a clean size-controlled claim. Phi-4 led the benchmark itself (best $\kappa$ and F1 of the eight;
Table~\ref{tab:bench}), and because a hate census penalizes missed positives most, we prioritized recall and
balance, on which it again led the deployed-label validation: accuracy 0.95, precision 0.91, recall 1.00
(Wilson 95\% CI [0.92, 1.00]), and $\kappa$=0.91 (Table~\ref{tab:val}). We therefore adopt Phi-4 as the
labeler of record. We validate it against two further models that bracket the precision--recall tradeoff:
DeepSeek-R1-Distill-32B is the \emph{highest-precision} labeler (P=0.95) but conservative ($R$=0.44), while
Qwen2.5 is more balanced ($\kappa$=0.59) yet, as an English-centric family, degrades sharply on Chinese
($\kappa$=0.40). No single model defines prevalence, so we report relative structure as primary and absolute
rates as calibrated ranges, and \emph{replicate every finding under a second independent labeler} (Qwen at
full corpus scale, while DeepSeek's reasoning latency precludes full-scale relabeling, so it serves as a
high-precision validation point). The validated pipeline yields 3{,}323 confirmed-hate comments. Contingency tests use $\chi^2$ with effect
size Cram\'er's $V$ and Benjamini--Hochberg FDR control, where for an $r\times c$ table of $N$ observations
\begin{equation}
V = \sqrt{\chi^2 \,/\, \bigl(N\,(\min(r,c)-1)\bigr)} .
\end{equation}
Resonance is modeled on Facebook with thread, length, and month controls and a within-thread comparison.
Moral-foundation comparisons are within-language, and the seed is fixed.

We validate both the binary hate gate and the characterization labels against human judgment. Because frame
and moral foundation are \emph{multi-label}, we score each audited item as correct ($n_y$), partial ($n_p$;
the model's label set overlaps but does not equal the human's), or wrong ($n_n$), with $N=n_y+n_p+n_n$, and
report an accuracy bracket
\begin{equation}
\mathrm{acc}_{\text{strict}} = \frac{n_y}{N}, \qquad
\mathrm{acc}_{\text{lenient}} = \frac{n_y+n_p}{N},
\end{equation}
treating a partial match as wrong (strict) or acceptable (lenient), with the true accuracy lying between.
On a human audit of $n$=30 confirmed-hate comments, the Phi-4 threat-frame label scores
0.83--0.90 (strict--lenient) and the moral-foundation label 0.80--0.97, with only 1 outright error in 30
(Table~\ref{tab:charact}), and per-language accuracy is $\ge$0.75. Frame and moral labels are thus moderately-to-well reliable, supporting the
\emph{relative} structural comparisons we report.

\begin{table}[tb]\centering\small
\caption{Labeler validation against human labels (independent random-stratified sample, $n$=89, best per
column in bold). Phi-4 (labeler of record) is the best balanced, DeepSeek is the highest-precision but most
conservative, and the spread is the precision--recall tradeoff that underlies the measurement caution.}
\label{tab:val}
\resizebox{\columnwidth}{!}{%
\begin{tabular}{lrrrr}
\toprule
Labeler & Acc & P & R & $\kappa$ \\
\midrule
\textbf{Phi-4 (record)} & \textbf{0.95} & 0.91 & \textbf{1.00} & \textbf{0.91} \\
\quad EN / ZH / MS & \multicolumn{4}{l}{$\kappa$ = 1.00 / 0.89 / 0.65} \\
DeepSeek-R1-Distill-32B & 0.72 & \textbf{0.95} & 0.44 & 0.43 \\
\quad EN / ZH / MS & \multicolumn{4}{l}{$\kappa$ = 0.38 / 0.38 / 0.56} \\
Qwen2.5 & 0.80 & 0.90 & 0.65 & 0.59 \\
\quad EN / ZH / MS & \multicolumn{4}{l}{$\kappa$ = 0.68 / 0.40 / 0.56} \\
\bottomrule
\end{tabular}}
\end{table}

\begin{table}[tb]\centering\small
\caption{Eight-model hate-gate benchmark vs the 211-item human gold (candidate-hate, hard slice), sorted by
$\kappa$ (best per column in bold). Model \emph{choice} dominates size, and agreement ceilings near 0.42. $^{\dagger}$4-bit.}
\label{tab:bench}
\resizebox{\columnwidth}{!}{%
\begin{tabular}{lrrrr}
\toprule
Model & P & R & F1 & $\kappa$ \\
\midrule
Phi-4 (14B) & 0.63 & 0.62 & \textbf{0.63} & \textbf{0.42} \\
DeepSeek-R1-Distill-Qwen-32B & \textbf{0.73} & 0.42 & 0.53 & 0.36 \\
Qwen2.5-32B & 0.58 & 0.55 & 0.56 & 0.32 \\
Qwen2.5-7B & 0.57 & 0.51 & 0.53 & 0.29 \\
Qwen2.5-14B & 0.52 & 0.65 & 0.58 & 0.28 \\
Qwen2.5-72B$^{\dagger}$ & 0.61 & 0.36 & 0.46 & 0.25 \\
Mistral-Small-24B & 0.70 & 0.25 & 0.37 & 0.22 \\
Gemma-2-9B & 0.43 & \textbf{0.84} & 0.57 & 0.17 \\
\bottomrule
\end{tabular}}
\end{table}

\begin{table}[tb]\centering\small
\caption{Characterization-label validation against human judgment ($n$=30). Strict
treats a partial multi-label match as wrong, lenient as acceptable.}
\label{tab:charact}
\resizebox{\columnwidth}{!}{%
\begin{tabular}{lrrr}
\toprule
Label & $n$ & strict & lenient \\
\midrule
Threat frame (RQ2) & 30 & 0.83 & 0.90 \\
Moral foundation (RQ3) & 30 & 0.80 & 0.97 \\
\bottomrule
\end{tabular}}
\end{table}

\paragraph{Independent lexicon cross-validation.} As a non-LLM check on the two characterization axes that
most shape our reading, moral foundation and affect, we re-score the English confirmed-hate comments with
fixed dictionaries. The Moral Foundations Dictionary \citep[MFD~2.0;][]{graham2009mft} places
\emph{sanctity/degradation} at the top of the five foundations, as do the LLM labels
(Table~\ref{tab:lexicon}); agreement on the full ordering is only moderate (Spearman $\rho$=0.40) and the
lexicon flattens the binding-over-individualizing gap, because a surface-word dictionary misses framing the
model infers (hate invoking loyalty or betrayal rarely uses dictionary loyalty words). For affect, LIWC2015
\citep{pennebaker2015liwc} reproduces the anger-dominant, fear-rare profile: anger words appear in 30\% of
comments versus anxiety in 14\% (a 2.2$\times$ gap mirroring the LLM's 62\% vs 10\%), against a strongly
negative tone (negative-emotion words in 53\% of comments, positive in 48\%). LIWC has no contempt or disgust
category, so those labels are not directly testable. Both lexicons recover the strongest patterns while
agreeing only moderately on absolute rates, reinforcing our decision to report relative structure as primary.

\begin{table}[tb]\centering\small
\caption{Independent lexicon cross-validation (English confirmed-hate, $n$=1{,}576). The Moral Foundations
Dictionary recovers the sanctity-first ordering of the LLM moral-foundation labels (\% of comments invoking
each foundation).}
\label{tab:lexicon}
\resizebox{\columnwidth}{!}{%
\begin{tabular}{lrr}
\toprule
Moral foundation & LLM label \% & MFD \% \\
\midrule
Sanctity & 55 & 24 \\
Loyalty & 51 & 22 \\
Authority & 27 & 16 \\
Fairness & 23 & 15 \\
Care & 16 & 23 \\
\bottomrule
\end{tabular}}
\end{table}

\section{Results}
\subsection*{RQ1: Is hate culturally contingent or cross-lingually shared?}
\begin{figure*}[t]\centering
\includegraphics[width=0.98\textwidth]{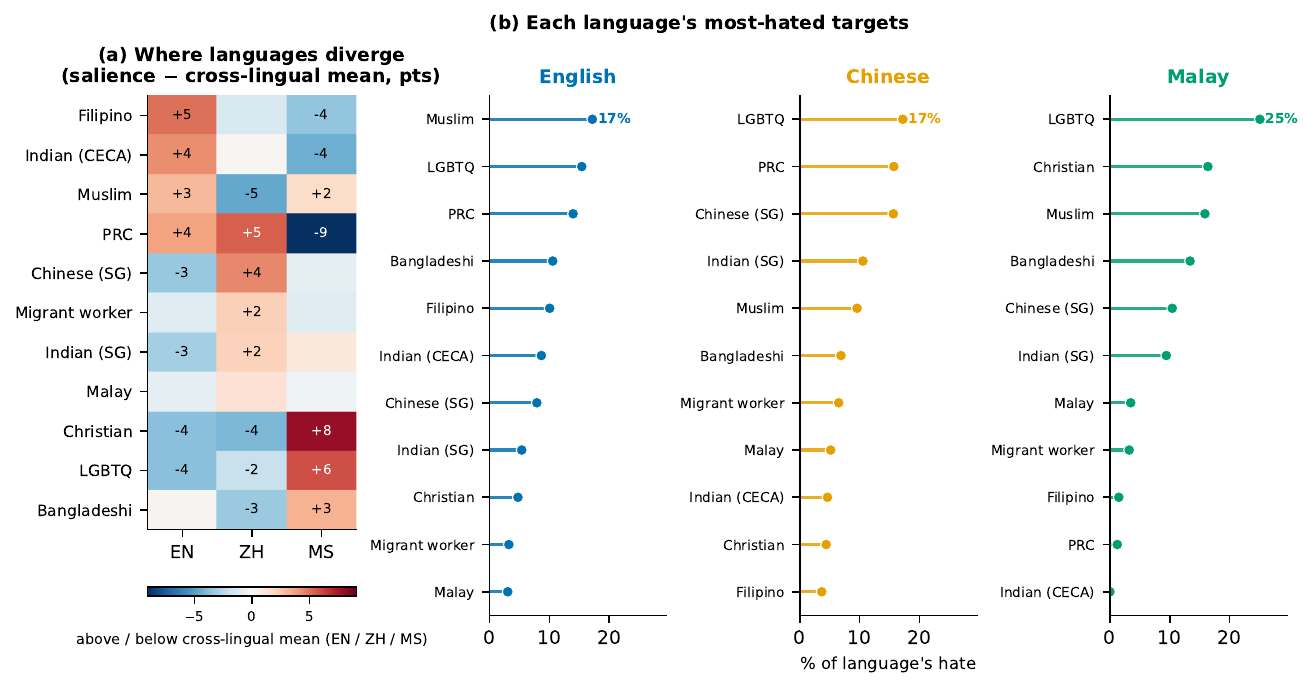}
\caption{Target salience is language-specific. \textbf{(a)} Divergence of each language from the cross-lingual
mean salience per target (percentage points): targets that are uniformly salient wash to neutral, while the
culturally divergent ones stand out---English over-indexes on foreign nationals, Chinese on the PRC/local-Chinese
axis, and Malay on religion and LGBTQ. \textbf{(b)} Each language's targets ranked by share of its
confirmed-hate. English is broad, Chinese centers the local-Chinese/PRC axis, and Malay centers religion
and LGBTQ.}
\label{fig:salience}
\end{figure*}

\paragraph{RQ1a (Targets): salience is cultural.} Which out-group a community fixates on differs systematically by
language (confirmed-hate volume; $\chi^2$=402, $V$=0.25). English discourse is broad, led by Muslim, PRC, and
LGBTQ targets. Chinese discourse centers the Chinese-Singaporean--mainland-PRC axis, and Malay discourse
centers religion and LGBTQ. The pattern is near-identical under both labelers (Figure~\ref{fig:salience}).

\begin{figure}[tb]\centering
\includegraphics[width=\columnwidth]{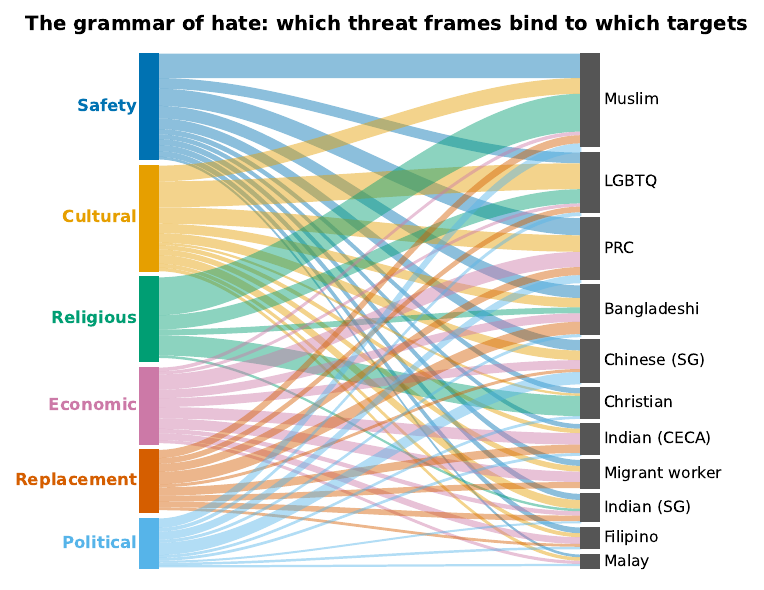}
\caption{The grammar of hate: an alluvial map of threat frame (left) to target group (right), ribbon width
$\propto$ co-occurrence. Religious framing binds to religious groups, material framing
(economic/safety/replacement) to migrant and foreign-national targets, and political-power to local-Chinese.}
\label{fig:frames}
\end{figure}

\paragraph{RQ1b (Frames): framing is structural.} Among hateful comments, threat-frame depends strongly on target group
($\chi^2$=1909, $V$=0.21; Table~\ref{tab:frames}) and the structure is theory-clean and robust across labelers.
Religious groups draw religious threat (Christian 63\%, Muslim 38\%), local-Chinese targets draw
political-power framing (27\%), and migrant and foreign-national targets draw \emph{material} threats, a mix of
economic (CECA 36\%, migrant-worker 35\%), safety/crime, and demographic-replacement framing. Cross-lingual
frame invariance is largely stable, with anti-Muslim hostility the most invariant (Figure~\ref{fig:frames}).

\begin{table}[tb]\centering\small
\caption{Threat-frame prevalence among hateful comments, by target group (Phi-4). Each row is the
percentage of that target's hate invoking each threat frame (bold = dominant frame). Rows are grouped by
dominant frame and columns ordered to expose the target$\to$frame mapping: migrant/foreign-national targets
draw economic threat, PRC safety, local-Chinese political-power, religious groups religious threat.}
\label{tab:frames}
\setlength{\tabcolsep}{4pt}
\resizebox{\columnwidth}{!}{%
\begin{tabular}{lrrrrrr}
\toprule
Target & \rotatebox{55}{Economic} & \rotatebox{55}{Safety/crime} & \rotatebox{55}{Political power}
 & \rotatebox{55}{Religious} & \rotatebox{55}{Cultural} & \rotatebox{55}{Replacement} \\
\midrule
CECA-Indian & \textbf{36\%} & 15\% & 9\% & 0\% & 14\% & 26\% \\
Migrant worker & \textbf{35\%} & 20\% & 3\% & 2\% & 19\% & 20\% \\
PRC & 23\% & \textbf{26\%} & 12\% & 0\% & 25\% & 12\% \\
Chinese-SG & 19\% & 23\% & \textbf{27\%} & 1\% & 21\% & 7\% \\
Muslim & 4\% & 24\% & 9\% & \textbf{38\%} & 16\% & 8\% \\
Christian & 2\% & 16\% & 9\% & \textbf{63\%} & 7\% & 1\% \\
\bottomrule
\end{tabular}}
\end{table}

\begin{figure}[tb]\centering
\includegraphics[width=\columnwidth]{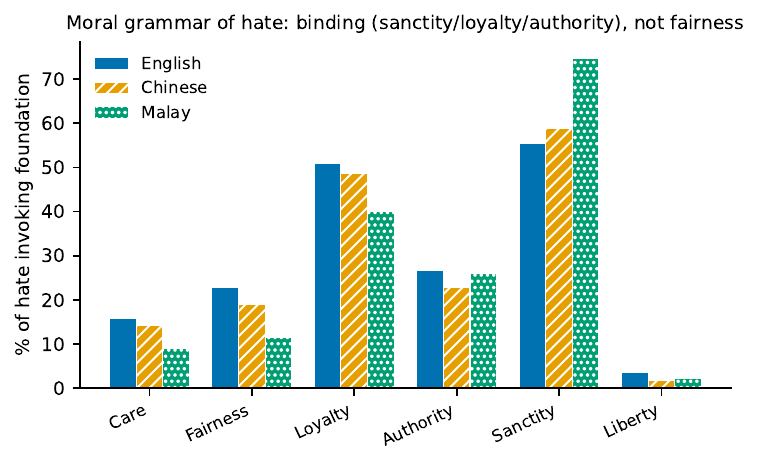}
\caption{The moral grammar of hate is binding-based (sanctity/loyalty/authority), not fairness, and the
profile is near-identical across the three language publics.}
\label{fig:mft}
\end{figure}

\paragraph{RQ1c (Moral grounding): near-universal and binding-based.} Across all three languages, the dominant
moral foundation of hate is \emph{sanctity/degradation} (English 55\%, Chinese 59\%, Malay 75\%),
followed by loyalty/betrayal and authority/subversion, the \emph{binding} foundations. Fairness/cheating is
low ($\le$23\%), and the pattern is confirmed by both labelers: hate's moral grammar is one of purity and
group loyalty (Figure~\ref{fig:mft}). This moral profile is also the \emph{least} cross-lingually variable axis
($V$=0.08; Figure~\ref{fig:hero}): the moral grammar of hate is shared even where its targets are not.

\subsection*{RQ2: How is hate substantively expressed?}
\paragraph{RQ2a (Narratives): which Singapore agendas carry hate.} Narrative prevalence differs by language ($\chi^2$=268, $V$=0.19;
Table~\ref{tab:expression}).
Anti-PRC sentiment is prominent in English (15\%) and Chinese (16\%), CMIO-racial framing is strongest in
Chinese (23\%), and CECA-conspiracy framing is largely English (11\%).

\paragraph{RQ2b (Affect): contempt-driven, with dehumanization.} Hate is contempt-driven (English 94\%, Chinese 96\%, Malay
96\%) and angry (56--62\%), with disgust (28--40\%) and rare fear: it looks down on rather than fears its
target. Dehumanization is present and non-trivial (21--35\%, animalistic and mechanistic).

\paragraph{RQ2c (Stance): an out-group, not anti-system, grievance.} Of confirmed hate, 31\% carries an
explicit anti-immigration stance (anti-establishment only 5\%), cleanly target-driven (migrant-worker 99\%,
CECA 98\%, PRC 59\%, versus LGBTQ 2\%, Christian 0\%). The hypothesized coupling of anti-immigration with
anti-establishment is rejected: they are mildly negatively associated (odds ratio 0.59, $p$=0.004).

\begin{table}[tb]\centering\small
\caption{How hate is expressed, by language (\% of each language's confirmed hate; multi-label, so columns need not sum to 100). Narratives differ by language ($\chi^2$=268, $V$=0.19); affect is contempt-driven across all three; stance is an out-group (anti-immigration) rather than anti-system grievance.}
\label{tab:expression}
\resizebox{\columnwidth}{!}{%
\begin{tabular}{lccc}
\toprule
\% of hate & EN & ZH & MS \\
\midrule
\multicolumn{4}{l}{\textit{Narrative (RQ2a)}} \\
\quad Anti-PRC & 15 & 16 & 4 \\
\quad CMIO-racial & 12 & 23 & 12 \\
\quad CECA-conspiracy & 11 & 10 & 3 \\
\quad Religious tension & 2 & 6 & 8 \\
\quad Migrant-dorm stigma & 7 & 10 & 3 \\
\addlinespace
\multicolumn{4}{l}{\textit{Affect (RQ2b)}} \\
\quad Contempt & 94 & 96 & 96 \\
\quad Anger & 62 & 56 & 56 \\
\quad Disgust & 28 & 35 & 40 \\
\quad Fear & 10 & 5 & 5 \\
\quad Dehumanization & 21 & 35 & 29 \\
\addlinespace
\multicolumn{4}{l}{\textit{Stance (RQ2c)}} \\
\quad Anti-immigration & 35 & 30 & 13 \\
\quad Anti-establishment & 7 & 4 & 3 \\
\bottomrule
\multicolumn{4}{l}{\footnotesize $n$ = 1584 EN / 1336 ZH / 403 MS confirmed-hate comments.} \\
\end{tabular}}
\end{table}

\subsection*{RQ3: Does the audience amplify what is produced?}
\begin{figure}[tb]\centering
\includegraphics[width=\columnwidth]{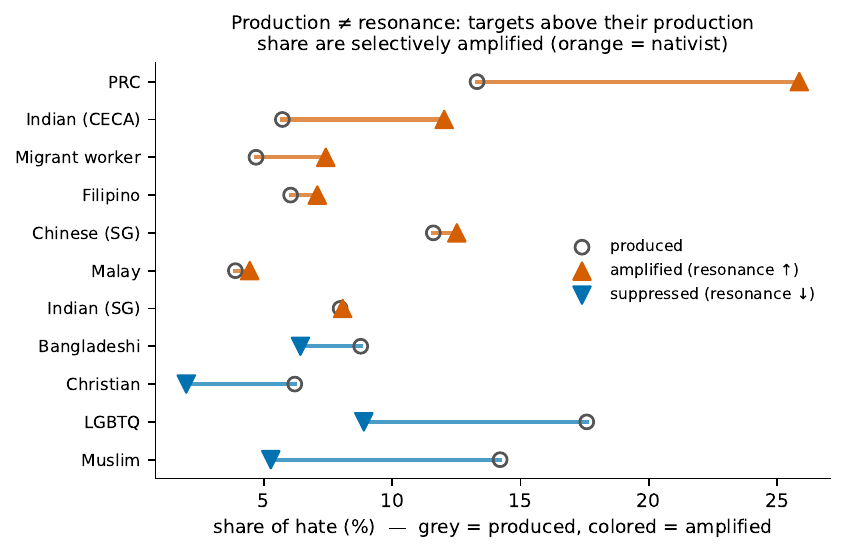}
\caption{Production $\neq$ resonance: for each target, share of hate \emph{produced} (grey) vs share
\emph{amplified} (received engagement, colored). Targets whose amplified share exceeds their produced share
are selectively amplified. Migrant/foreign-national (nativist) targets (orange) gain, while religious/identity
targets (blue) lose. Sorted by amplification gap.}
\label{fig:resonance}
\end{figure}

\paragraph{RQ3a (Resonance): selectively nativist.} On Facebook, hateful comments are amplified \emph{less}
than neutral target-mentions. The dissociation survives the strongest available confound control: in a
\emph{within-thread} paired comparison over 1{,}033 threads that contain both hate and neutral
target-mentions (holding thread, space, topic, and time fixed), hate is under-amplified relative to neutral
content in the same thread (mean difference $-0.056$ in any-engagement rate; Wilcoxon $p$=3.1$\times$10$^{-4}$).
Among hate, amplification is selectively nativist (Figure~\ref{fig:resonance}). A logistic model controlling
for target, language, log thread-size, log comment-length, and month (FB hate, $n$=3{,}111; McFadden
$R^2$=0.14) confirms the ranking with 95\% CIs: anti-CECA $+1.71$ [1.28, 2.13], anti-PRC $+1.52$ [1.16, 1.87],
and migrant-worker $+1.24$ [0.77, 1.71] are all significantly amplified, whereas religious targets sit at the
floor (Muslim $-0.28$, Christian $-0.37$, both n.s.). What a community produces dissociates from what it
amplifies, and what it amplifies is anti-immigrant. We measure \emph{resonance} (reach), not endorsement:
engagement counts cannot separate approving from oppositional reactions, so a comment drawing counter-speech
also accrues engagement. The selective-amplification result is therefore a reach-level association net of the
controls above. The pattern is consistent with out-group animosity and moral-emotional content driving
engagement \citep{rathje2021outgroup,brady2017contagion,brady2021amplify}.

\begin{figure}[tb]\centering
\includegraphics[width=\columnwidth]{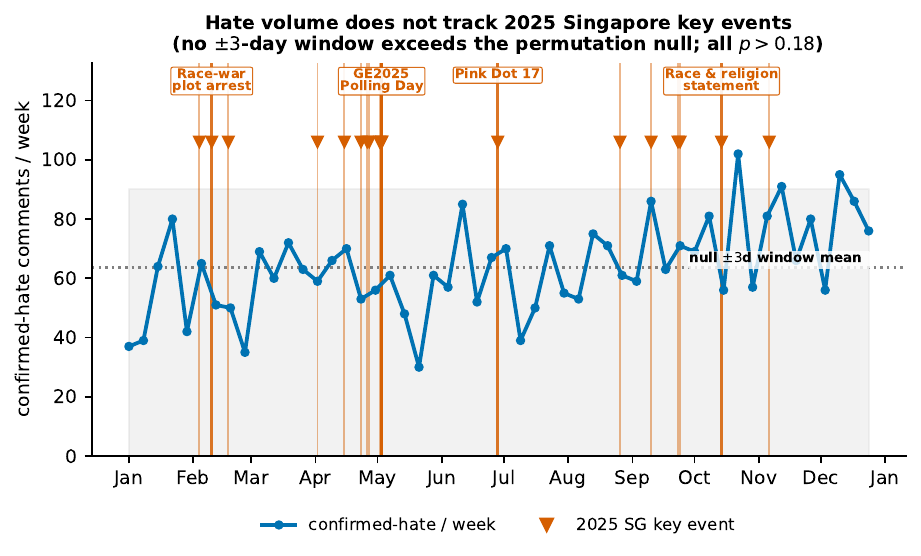}
\caption{Weekly confirmed-hate volume across 2025 (blue) against the 2025 Singapore key-events calendar (orange).
No $\pm3$-day event window exceeds the permutation null (dotted line, mean window count; shading marks its
95th-percentile range); all event $p>0.18$. Hate volume is steady and event-insensitive at this scale.}
\label{fig:temporal}
\end{figure}

\paragraph{RQ3b (Temporal): hate volume does not track 2025 Singapore key events.} In a $\pm3$-day window against a
permutation null, confirmed-hate volume shows no significant spike around any 2025 Singapore key event
(general-election milestones, security detentions, racial or religious incidents, all $p>0.18$;
Figure~\ref{fig:temporal}). The pattern holds even for the events one would most expect to mobilize
out-group hostility: the February detention of a teenager who had planned a Chinese-versus-Malay ``race
war,'' the May general-election polling day (whose campaign foregrounded immigration and the CECA trade
pact), the June Pink Dot LGBTQ rally, and the October ministerial statement on race and religion each
sit on an unremarkable week. At the
confirmed-hate sample size (about 64 comments per $\pm3$-day window) the test has 80\% power only for
surges of roughly $+53\%$ (about 34 extra comments per window), so we read this as no evidence of
event-driven spikes rather than evidence of a flat baseline. To check that the pooled volume is not
masking a target-specific reaction, we also ran a per-target event study: each event is mapped to the
out-group it concerns, and that slice is tested against its own permutation null
(Table~\ref{tab:eventstudy}). No event-implicated slice exceeds its null ($p>0.05$ throughout); the closest
is LGBTQ-directed hate around the Pink Dot rally (roughly double the null, $p$=0.06), which still falls short
and would not survive correction across the ten events.

\begin{table}[tb]\centering\small
\caption{Per-target event study: for each 2025 event, the $\pm3$-day window count of confirmed-hate \emph{against the out-group the event concerns}, versus that target's own permutation null (10{,}000 random windows). No event-implicated slice exceeds its null ($p>0.05$ throughout), so the pooled null (Figure~\ref{fig:temporal}) is not masking a target-specific spike (rows ordered by $p$).}
\label{tab:eventstudy}
\resizebox{\columnwidth}{!}{%
\begin{tabular}{llrrr}
\toprule
Event & Target slice & Obs & Null & $p$ \\
\midrule
Pink Dot 17 & LGBTQ & 23 & 11 & 0.06 \\
Race \& religion statement & Malay/Muslim/Chinese/Indian & 35 & 24 & 0.13 \\
CECA falsehoods rebuttal & CECA & 7 & 4 & 0.14 \\
GE2025 Polling Day (immigration) & CECA/Migrant/PRC & 21 & 15 & 0.14 \\
Election forum (immigration/CECA) & CECA/Migrant/PRC & 17 & 15 & 0.38 \\
ISD race-war plot arrest & Malay/Muslim/Chinese & 18 & 19 & 0.52 \\
Intra-CMIO TikTok ill-will & Chinese/Malay/Indian & 13 & 15 & 0.59 \\
Mosque pork-parcel incident & Muslim & 6 & 9 & 0.77 \\
Asatizah/communal-voting row & Muslim/Malay & 8 & 11 & 0.80 \\
Mosque-attack plot arrest & Muslim/Malay & 7 & 11 & 0.84 \\
\bottomrule
\end{tabular}}
\end{table}

\section{Discussion}
The structure of online hate decomposes into axes that do not move together, and the decomposition is
\emph{ordered}: cross-lingual divergence falls monotonically from target ($V$=0.25) to frame (0.15) to
moral foundation (0.08) to emotion (0.07) (Figure~\ref{fig:hero}). This is the paper's central claim, a
layered cultural contingency in which the \emph{what} of hate is locally selected while the \emph{how} and
\emph{why} are a shared cross-lingual grammar. It goes beyond confirming that Integrated Threat Theory or
Moral Foundations Theory ``apply'': it locates \emph{which} components of intergroup hostility are
culturally plastic and which are near-invariant, a distinction that a single-language study cannot make and
that a cross-lingual, single-society design is built to. The audience layer adds a fourth: what is amplified
(resonance) is selectively nativist and dissociated from what is produced. Treating ``online hate'' as a
single prevalence to be measured and reduced obscures this layered structure.

Three lenses jointly ground this architecture. Social Identity Theory supplies the universal engine (group
categorization and out-group derogation) while leaving the salient out-group locally determined
\citep{tajfel1979integrative}, explaining why the \emph{what} varies. The binding moral foundations
(sanctity, loyalty, authority), which are universal but differentially weighted across cultures
\citep{graham2011mapping,atari2020iran} and specifically tied to out-group hostility
\citep{kugler2014another}, supply the shared moral grammar. The disgust$\rightarrow$dehumanization
pathway \citep{hodson2007interpersonal,haslam2006dehumanization} fuses the sanctity foundation with
threat-based prejudice into one mechanism \citep{bianco2023binding} rather than two parallel lenses. The
combination predicts exactly what we observe: the categorization target is culturally plastic while the
moral-affective form of the resulting hate is cross-lingually conserved.

The salience--frame split has a practical reading for cross-lingual moderation. Because targets are
language-specific but the frame-to-target structure is largely shared, a detector or policy keyed on
\emph{frames} is a better transfer candidate than one keyed on \emph{targets}, which must be re-localized per
language public. The production--resonance dissociation reframes a moderation target the field leaves implicit:
the harm of hate is partly a function of its reach, and reach is a filter with its own preferences, not a
neutral amplifier of what is produced.

Beyond hate, the study surfaces a measurement caution for LLM-based content analysis. For a construct as
contested as hate, absolute prevalence is not well defined at the annotator level: eight capable models
ceiling at $\kappa\approx0.42$, and an English-centric family degrades on Chinese. The practical lesson is that
relative contrasts are the trustworthy unit, and absolute rates require human anchoring and should be reported
as calibrated ranges.

\section{Limitations}
Our findings hold within several scope conditions. The corpus is actor-less, so we characterize comments
rather than users and make no claim about who produces hate or who is vulnerable to it. Because absolute
hate prevalence is not well defined across annotators, we treat relative structure as the primary object,
report absolute rates as calibrated ranges, and compare moral foundations within rather than across
languages. The associations we report are real but modest in size ($V\approx0.2$), and we read them as such.
Our theoretical lenses are used descriptively, not causally, and no content is attributed to any person,
group, or state.

Three conditions most qualify the cross-lingual reading. First, Chinese and Malay are labeled on
machine-English translations, and a native-language audit finds substantial, if imperfect, agreement with the
translated-text labels (Cohen's $\kappa$=0.56), so cross-lingual comparisons rest on relative structure, and
Malay, the smallest cell, most warrants native-language replication. Second, the corpus is a crawl of a
curated set of (largely news, community, and alternative-news) Singapore sources rather than a random sample,
so it best represents news-anchored public discourse, and because Chinese and Malay content sits almost
entirely on Facebook, language is partly confounded with platform. Third, the binary hate gate, threat frame, and moral
foundation are validated against human judgment (Tables~\ref{tab:val}, \ref{tab:charact}), moral foundation
and affect are additionally corroborated by independent lexicons (Table~\ref{tab:lexicon}), and the narrative
and stance labels rest on cross-model replication alone. We flag the handful of patterns that
shift with the labeler and do not build on them. The study covers one society over one year, and its
findings should be read in that frame.

\section{Ethical Considerations}
The study is a secondary analysis of provider-released, actor-less data and profiles no individuals. We follow
a strict harm-minimization protocol: we store labels and item identifiers only, never raw hateful text or slur
lists, and all illustrative examples are paraphrased. The most severe content is diverted to an aggregate count
and never characterized. We name no person, organization, platform, state, or community as responsible, and
``communities'' denote language strata described neutrally. Provider-supplied affect scores are used only as a
discovery signal and are not audited. The methods characterize hate to inform moderation and counter-speech and
are not suitable as a deployable targeting instrument. The relative-only, actor-less, no-attribution design is
deliberate on this point.

\section*{Mandatory Statements}
\textbf{Data Availability.} Derived labels and code are available from the author; raw provider data is
governed by the data-sharing agreement. \textbf{Author Contributions (CRediT).} E.~Ferrara is the sole author
(all CRediT roles). \textbf{Conflict of Interest.} The author declares no competing interests.
\textbf{Acknowledgments.} The author thanks the CATOS Resonance team and A*STAR collaborators for the data
corpus. \textbf{Funding.} Supported by NSF HCC award \#2331722. \textbf{AI-Usage Disclosure.} Large language
models served as labeling instruments (Methods), benchmarked against human gold. AI assistance in drafting is
disclosed per venue policy.

\balance
\newpage
\balance
\bibliographystyle{plainnat}
\bibliography{refs}

@incollection{stephan2000itt,
  author = {Stephan, Walter G. and Stephan, Cookie White},
  title = {An Integrated Threat Theory of Prejudice},
  booktitle = {Reducing Prejudice and Discrimination},
  editor = {Oskamp, Stuart},
  pages = {23--45}, year = {2000}, publisher = {Lawrence Erlbaum},
  doi = {10.4324/9781410605634-7}
}

@incollection{stephan2016itt,
  author = {Stephan, Walter G. and Stephan, Cookie White},
  title = {Intergroup Threats},
  booktitle = {The Cambridge Handbook of the Psychology of Prejudice},
  editor = {Sibley, Chris G. and Barlow, Fiona Kate},
  pages = {131--148}, year = {2016}, publisher = {Cambridge University Press},
  doi = {10.1017/9781316161579.007}
}

@article{graham2009mft,
  author = {Graham, Jesse and Haidt, Jonathan and Nosek, Brian A.},
  title = {Liberals and Conservatives Rely on Different Sets of Moral Foundations},
  journal = {Journal of Personality and Social Psychology},
  volume = {96}, number = {5}, pages = {1029--1046}, year = {2009},
  doi = {10.1037/a0015141}
}

@article{hopp2021emfd,
  author = {Hopp, Frederic R. and Fisher, Jacob T. and Cornell, Devin and Huskey, Richard and Weber, Ren{\'e}},
  title = {The Extended Moral Foundations Dictionary ({eMFD}): Development and Applications of a Crowd-Sourced Approach to Extracting Moral Intuitions from Text},
  journal = {Behavior Research Methods},
  volume = {53}, pages = {232--246}, year = {2021},
  doi = {10.3758/s13428-020-01433-0}
}

@inproceedings{jiang2025moral,
  author = {Jiang, Julie and Luceri, Luca and Ferrara, Emilio},
  title = {Moral Values Underpinning {COVID-19} Online Communication Patterns},
  booktitle = {Companion Proceedings of the ACM Web Conference 2025 (WWW '25 Companion)},
  year = {2025}, doi = {10.1145/3701716.3717538}
}

@misc{jiang2024toxicity,
  author = {Jiang, Julie and Luceri, Luca and Walther, Joseph B. and Ferrara, Emilio},
  title = {Social Approval and Network Homophily as Motivators of Online Toxicity},
  year = {2024}, eprint = {2310.07779}, archivePrefix = {arXiv},
  howpublished = {arXiv preprint arXiv:2310.07779}
}

@inproceedings{ng2024sghatecheck,
  author = {Ng, Ri Chi and Prakash, Nirmalendu and Hee, Ming Shan and Choo, Kenny Tsu Wei and Lee, Roy Ka-Wei},
  title = {{SGHateCheck}: Functional Tests for Detecting Hate Speech in Low-Resource Languages of Singapore},
  booktitle = {Proceedings of the 8th Workshop on Online Abuse and Harms (WOAH 2024)},
  pages = {312--331}, year = {2024}, doi = {10.18653/v1/2024.woah-1.24}
}

@inproceedings{haider2023lowresource,
  author = {Haider, Samar and Luceri, Luca and Deb, Ashok and Badawy, Adam and Peng, Nanyun and Ferrara, Emilio},
  title = {Detecting Social Media Manipulation in Low-Resource Languages},
  booktitle = {Companion Proceedings of the ACM Web Conference 2023 (WWW '23 Companion)},
  year = {2023}, doi = {10.1145/3543873.3587615}
}

@inproceedings{dey2025cultural,
  author = {Dey, Priyanka and Bothra, Aayush and Khanter, Yugal and Zhao, Jieyu and Ferrara, Emilio},
  title = {Can {LLMs} Express Personality Across Cultures? Introducing {CulturalPersonas} for Evaluating Trait Alignment},
  booktitle = {Findings of the Association for Computational Linguistics: EMNLP 2025},
  year = {2025}, doi = {10.18653/v1/2025.findings-emnlp.1101}
}

@article{lerman2025moderation,
  author = {Lerman, Kristina and Chu, Minh Duc and Bickham, Charles and Luceri, Luca and Ferrara, Emilio},
  title = {Safe spaces or toxic places? Content moderation and social dynamics of online eating disorder communities},
  journal = {EPJ Data Science}, volume = {14}, number = {1}, year = {2025},
  doi = {10.1140/epjds/s13688-025-00575-5}
}

@techreport{pennebaker2015liwc,
  author = {Pennebaker, James W. and Boyd, Ryan L. and Jordan, Kayla and Blackburn, Kate},
  title = {The Development and Psychometric Properties of {LIWC2015}},
  institution = {University of Texas at Austin}, address = {Austin, TX}, year = {2015}
}

@inproceedings{deng2022cold,
  author = {Deng, Jiawen and Zhou, Jingyan and Sun, Hao and Zheng, Chujie and Mi, Fei and Meng, Helen and Huang, Minlie},
  title = {{COLD}: A Benchmark for Chinese Offensive Language Detection},
  booktitle = {Proceedings of EMNLP 2022},
  pages = {11580--11599}, year = {2022}, doi = {10.18653/v1/2022.emnlp-main.796}
}

@inproceedings{lu2023toxicn,
  author = {Lu, Junyu and Xu, Bo and Zhang, Xiaokun and Min, Changrong and Yang, Liang and Lin, Hongfei},
  title = {Facilitating Fine-Grained Detection of Chinese Toxic Language: Hierarchical Taxonomy, Resources, and Benchmarks},
  booktitle = {Proceedings of ACL 2023 (Long Papers)},
  pages = {16235--16250}, year = {2023}, doi = {10.18653/v1/2023.acl-long.898}
}

@inproceedings{ibrohim2019indonesian,
  author = {Ibrohim, Muhammad Okky and Budi, Indra},
  title = {Multi-label Hate Speech and Abusive Language Detection in Indonesian Twitter},
  booktitle = {Proceedings of the Third Workshop on Abusive Language Online (ALW3)},
  pages = {46--57}, year = {2019}, doi = {10.18653/v1/W19-3506}
}

@inproceedings{rottger2021hatecheck,
  author = {R{\"o}ttger, Paul and Vidgen, Bertie and Nguyen, Dong and Waseem, Zeerak and Margetts, Helen and Pierrehumbert, Janet},
  title = {{HateCheck}: Functional Tests for Hate Speech Detection Models},
  booktitle = {Proceedings of ACL-IJCNLP 2021 (Long Papers)},
  pages = {41--58}, year = {2021}, doi = {10.18653/v1/2021.acl-long.4}
}

@inproceedings{rottger2022multihatecheck,
  author = {R{\"o}ttger, Paul and Seelawi, Haitham and Nozza, Debora and Talat, Zeerak and Vidgen, Bertie},
  title = {Multilingual {HateCheck}: Functional Tests for Multilingual Hate Speech Detection Models},
  booktitle = {Proceedings of WOAH 2022}, pages = {154--169}, year = {2022},
  doi = {10.18653/v1/2022.woah-1.15}
}

@article{brady2017contagion,
  author = {Brady, William J. and Wills, Julian A. and Jost, John T. and Tucker, Joshua A. and Van Bavel, Jay J.},
  title = {Emotion Shapes the Diffusion of Moralized Content in Social Networks},
  journal = {Proceedings of the National Academy of Sciences},
  volume = {114}, number = {28}, pages = {7313--7318}, year = {2017},
  doi = {10.1073/pnas.1618923114}
}

@article{brady2021amplify,
  author = {Brady, William J. and McLoughlin, Killian and Doan, Tuan N. and Crockett, Molly J.},
  title = {How Social Learning Amplifies Moral Outrage Expression in Online Social Networks},
  journal = {Science Advances}, volume = {7}, number = {33}, pages = {eabe5641}, year = {2021},
  doi = {10.1126/sciadv.abe5641}
}

@inproceedings{mathew2019counterspeech,
  author = {Mathew, Binny and Saha, Punyajoy and Tharad, Hardik and Rajgaria, Subham and Singhania, Prajwal and Maity, Suman Kalyan and Goyal, Pawan and Mukherjee, Animesh},
  title = {Thou Shalt Not Hate: Countering Online Hate Speech},
  booktitle = {Proceedings of ICWSM 2019}, volume = {13}, number = {1}, pages = {369--380}, year = {2019},
  doi = {10.1609/icwsm.v13i01.3237}
}

@article{garland2022counterspeech,
  author = {Garland, Joshua and Ghazi-Zahedi, Keyan and Young, Jean-Gabriel and H{\'e}bert-Dufresne, Laurent and Galesic, Mirta},
  title = {Impact and Dynamics of Hate and Counter Speech Online},
  journal = {EPJ Data Science}, volume = {11}, pages = {3}, year = {2022},
  doi = {10.1140/epjds/s13688-021-00314-6}
}

@article{herold2025replacement,
  author = {Herold, Maik},
  title = {Who Believes in the ``Great Replacement''? Political Attitudes and Democratic Alienation Among Supporters of Immigration-Related Conspiracy Theories in Europe},
  journal = {Social Science Quarterly}, year = {2025}, doi = {10.1111/ssqu.13481}
}

@article{gilardi2023chatgpt,
  author = {Gilardi, Fabrizio and Alizadeh, Meysam and Kubli, Ma{\"e}l},
  title = {{ChatGPT} Outperforms Crowd Workers for Text-Annotation Tasks},
  journal = {Proceedings of the National Academy of Sciences},
  volume = {120}, number = {30}, pages = {e2305016120}, year = {2023},
  doi = {10.1073/pnas.2305016120}
}

@article{johnson2019hate,
  author = {Johnson, N. F. and Leahy, R. and Restrepo, N. J. and Velasquez, N. and Zheng, M. and Manrique, P. and Devkota, P. and Wuchty, S.},
  title = {Hidden Resilience and Adaptive Dynamics of the Global Online Hate Ecology},
  journal = {Nature}, volume = {573}, number = {7773}, pages = {261--265}, year = {2019},
  doi = {10.1038/s41586-019-1494-7}}

@inproceedings{davidson2017automated,
  author = {Davidson, Thomas and Warmsley, Dana and Macy, Michael and Weber, Ingmar},
  title = {Automated Hate Speech Detection and the Problem of Offensive Language},
  booktitle = {Proceedings of ICWSM 2017}, volume = {11}, number = {1}, pages = {512--515}, year = {2017},
  doi = {10.1609/icwsm.v11i1.14955}}

@inproceedings{elsherief2021latent,
  author = {ElSherief, Mai and Ziems, Caleb and Nguyen, David and Roy, Soumya and Wang, Diba and Yang, Diyi},
  title = {Latent Hatred: A Benchmark for Understanding Implicit Hate Speech},
  booktitle = {Proceedings of EMNLP 2021}, pages = {345--363}, year = {2021},
  doi = {10.18653/v1/2021.emnlp-main.29}}

@inproceedings{mendelsohn2021framing,
  author = {Mendelsohn, Julia and Budak, Ceren and Jurgens, David},
  title = {Modeling Framing in Immigration Discourse on Social Media},
  booktitle = {Proceedings of NAACL 2021}, pages = {2219--2263}, year = {2021},
  doi = {10.18653/v1/2021.naacl-main.179}}

@article{rathje2021outgroup,
  author = {Rathje, Steve and Van Bavel, Jay J. and van der Linden, Sander},
  title = {Out-group Animosity Drives Engagement on Social Media},
  journal = {Proceedings of the National Academy of Sciences}, volume = {118}, number = {26}, pages = {e2024292118}, year = {2021},
  doi = {10.1073/pnas.2024292118}}

@article{solovev2022moralized,
  author = {Solovev, Kirill and Pr{\"o}llochs, Nicolas},
  title = {Moralized Language Predicts Hate Speech on Social Media},
  journal = {PNAS Nexus}, volume = {1}, number = {5}, pages = {pgac281}, year = {2022},
  doi = {10.1093/pnasnexus/pgac281}}

@article{hangartner2021empathy,
  author = {Hangartner, Dominik and Gennaro, Gloria and Alasiri, Sary and others},
  title = {Empathy-based Counterspeech Can Reduce Racist Hate Speech in a Social Media Field Experiment},
  journal = {Proceedings of the National Academy of Sciences}, volume = {118}, number = {50}, pages = {e2116310118}, year = {2021},
  doi = {10.1073/pnas.2116310118}}

@inproceedings{vidgen2020eastasian,
  author = {Vidgen, Bertie and Botelho, Austin and Broniatowski, David and Guest, Ella and Hall, Matthew and Margetts, Helen and Tromble, Rebekah and Waseem, Zeerak and Hale, Scott},
  title = {Detecting East Asian Prejudice on Social Media},
  booktitle = {Proceedings of the 4th Workshop on Online Abuse and Harms (WOAH)}, pages = {162--172}, year = {2020},
  doi = {10.18653/v1/2020.alw-1.19}}

@inproceedings{sap2022annotators,
  author = {Sap, Maarten and Swayamdipta, Swabha and Vianna, Laura and Zhou, Xuhui and Choi, Yejin and Smith, Noah A.},
  title = {Annotators with Attitudes: How Annotator Beliefs and Identities Bias Toxic Language Detection},
  booktitle = {Proceedings of NAACL 2022}, pages = {5884--5906}, year = {2022},
  doi = {10.18653/v1/2022.naacl-main.431}}

@article{davani2022disagreements,
  author = {Davani, Aida Mostafazadeh and D{\'i}az, Mark and Prabhakaran, Vinodkumar},
  title = {Dealing with Disagreements: Looking Beyond the Majority Vote in Subjective Annotations},
  journal = {Transactions of the Association for Computational Linguistics}, volume = {10}, pages = {92--110}, year = {2022},
  doi = {10.1162/tacl_a_00449}}

@incollection{tajfel1979integrative,
  author = {Tajfel, Henri and Turner, John C.},
  title = {An Integrative Theory of Intergroup Conflict},
  booktitle = {The Social Psychology of Intergroup Relations}, pages = {33--47}, year = {1979}, publisher = {Brooks/Cole},
  doi = {10.4324/9780203505984-16}}

@article{graham2011mapping,
  author = {Graham, Jesse and Nosek, Brian A. and Haidt, Jonathan and Iyer, Ravi and Koleva, Spassena and Ditto, Peter H.},
  title = {Mapping the Moral Domain},
  journal = {Journal of Personality and Social Psychology}, volume = {101}, number = {2}, pages = {366--385}, year = {2011},
  doi = {10.1037/a0021847}}

@article{atari2020iran,
  author = {Atari, Mohammad and Graham, Jesse and Dehghani, Morteza},
  title = {Foundations of Morality in Iran},
  journal = {Evolution and Human Behavior}, volume = {41}, number = {5}, pages = {367--384}, year = {2020},
  doi = {10.1016/j.evolhumbehav.2020.07.014}}

@article{haslam2006dehumanization,
  author = {Haslam, Nick},
  title = {Dehumanization: An Integrative Review},
  journal = {Personality and Social Psychology Review}, volume = {10}, number = {3}, pages = {252--264}, year = {2006},
  doi = {10.1207/s15327957pspr1003_4}}

@article{hodson2007interpersonal,
  author = {Hodson, Gordon and Costello, Kimberly},
  title = {Interpersonal Disgust, Ideological Orientations, and Dehumanization as Predictors of Intergroup Attitudes},
  journal = {Psychological Science}, volume = {18}, number = {8}, pages = {691--698}, year = {2007},
  doi = {10.1111/j.1467-9280.2007.01962.x}}

@article{bianco2023binding,
  author = {Bianco, Federica and Kosic, Ankica},
  title = {The Effects of Binding Moral Foundations on Prejudiced Attitudes toward Migrants: The Mediation Role of Perceived Realistic and Symbolic Threats},
  journal = {Genealogy}, volume = {7}, number = {3}, pages = {65}, year = {2023},
  doi = {10.3390/genealogy7030065}}

@article{kugler2014another,
  author = {Kugler, Matthew and Jost, John T. and Noorbaloochi, Sharareh},
  title = {Another Look at Moral Foundations Theory: Do Authoritarianism and Social Dominance Orientation Explain Liberal-Conservative Differences in ``Moral'' Intuitions?},
  journal = {Social Justice Research}, volume = {27}, number = {4}, pages = {413--431}, year = {2014},
  doi = {10.1007/s11211-014-0223-5}}

@article{mackie2000intergroup,
  author = {Mackie, Diane M. and Devos, Thierry and Smith, Eliot R.},
  title = {Intergroup Emotions: Explaining Offensive Action Tendencies in an Intergroup Context},
  journal = {Journal of Personality and Social Psychology}, volume = {79}, number = {4}, pages = {602--616}, year = {2000},
  doi = {10.1037/0022-3514.79.4.602}}

@article{opotow1990moral,
  author = {Opotow, Susan},
  title = {Moral Exclusion and Injustice: An Introduction},
  journal = {Journal of Social Issues}, volume = {46}, number = {1}, pages = {1--20}, year = {1990},
  doi = {10.1111/j.1540-4560.1990.tb00268.x}}
\end{document}